\documentclass{comjnl}

\usepackage{amsmath}
\usepackage{cite}
\usepackage{amsmath,amssymb,amsfonts}
\usepackage{algorithmic}
\usepackage{graphicx}
\usepackage{textcomp}
\usepackage{xcolor,pifont}
\usepackage{etoolbox,xstring,mfirstuc,textcase}

\usepackage{amsmath,graphicx,subfig,amsthm,mathrsfs,listings,amssymb,xcolor}
\usepackage{indentfirst,boxedminipage}

\usepackage[colorlinks,linkcolor=blue,anchorcolor=blue,citecolor=blue]{hyperref}

\begin{document}

\title[On the Modulus   in Matching Vector Codes]{On the Modulus   in Matching Vector Codes}
\author{Lin Zhu, Wen Ming Li, Liang Feng Zhang\thanks{corresponding author}}
\affiliation{School of Information Science and Technology,  ShanghaiTech University, \\ Shanghai 201210, China}
\email{zhanglf@shanghaitech.edu.cn (*corresponding author)}

\shortauthors{L. Zhu, W. Li, L.F. Zhang}


\begin{abstract}
A $k$-query locally decodable code (LDC)  $C$
allows one to encode any $n$-symbol  message $x$ as a codeword $C(x)$ of $N$ symbols such that each symbol
of $x$ can be   recovered by looking at $k$ symbols
of $C(x)$, even if a constant fraction of $C(x)$ have been corrupted.
Currently, the best known LDCs are matching vector codes (MVCs).
A  modulus $m=p_1^{\alpha_1}p_2^{\alpha_2}\cdots p_r^{\alpha_r}$ may result in an MVC with $k\leq 2^r$
and  $N=\exp(\exp(O((\log n)^{1-1/r} (\log\log n)^{1/r})))$.
The  $m$ is  {\em good} if it is possible to have
 $k<2^r$.
The good numbers yield more efficient
MVCs.
Prior to this work, there are only {\em finitely many} good numbers.
All of them were obtained via computer search and  have the form $m=p_1p_2$.
In this paper, we   study
good numbers of the form $m=p_1^{\alpha_1}p_2^{\alpha_2}$.
We show that  if $m=p_1^{\alpha_1}p_2^{\alpha_2}$ is good,
 then any multiple
 of $m$ of the form $p_1^{\beta_1}p_2^{\beta_2}$ must  be good as well. Given a good number
  $m=p_1^{\alpha_1}p_2^{\alpha_2}$, we   show an explicit method of  obtaining
    smaller good numbers that have the same   prime divisors.
Our  approach  yields {\em infinitely many} new good numbers.
\end{abstract}

\maketitle

\section{Introduction}

Classical error-correcting codes
allow one to encode any $n$-bit  message  $x$
as an $N$-bit codeword $C(x)$  such that   $x$ can still be recovered,
even if a constant fraction of $C(x)$  have been corrupted.
  The disadvantage of such codes is that one has to read all or most of
the codeword to recover any information about $x$.
As a better solution for decoding particular bits of the message, a   $(k,\delta,\epsilon)$-{\em locally decodable code} (LDC) \cite{KT00}
encodes any $n$-bit message $x$ to an $N$-bit codeword, such that
any message bit $x_i$ can be recovered with probability
$\geq 1-\epsilon$, by a randomized decoding procedure that
makes at most $k$ queries, even if $\delta N$ bits of $C(x)$
have been corrupted.
  Such codes have   interesting applications \cite{Gas04,Tre04} in cryptography and
complexity theory.
For an efficient   LDC,  both
 the code {\em length} $N$ and the {\em query complexity}
$k$ should be as small as possible,  as functions of $n$.

Following \cite{KT00,GKST06,Woo07},
Gasarch \cite{Gas04} and Goldreich \cite{GKST06}  conjectured that
for any constant $k$,
 the   length  $N$ of a $k$-query LDC should be  $\exp(n^{\Omega(1)})$. Yekhanin \cite{Yek08}
 disproved this conjecture with a 3-query LDC of
 length  $\exp(\exp(O(\log n/\log\log n)))$, assuming that there are
 infinitely many Mersenne primes.
For any   $r\geq 2$, Efremenko \cite{Efr09}  provided a construction of
$2^r$-query LDCs of length {$N_r=\exp(\exp(O((\log n)^{1-1/r} (\log\log n)^{1/r})))$} under no assumptions, and in particular a 3-query LDC   when $r=2$.
Such codes have been reformulated and called {\em matching vector codes} (MVCs) in \cite{DGY11}.

The MVCs in \cite{Efr09} are based on two ingredients: $S$-matching family and
$S$-decoding polynomial. For $r\geq 2$, let   $\mathcal{M}_r$ be the set of   integers of the form
$m=p_1^{\alpha_1}p_2^{\alpha_2}\cdots p_r^{\alpha_r}$, where
$p_1,p_2,\ldots,p_r>2$  are {distinct primes} and $\alpha_1,\alpha_2,\ldots,\alpha_r>0$.
 The existence of both ingredients in MVCs
  depends on a {\em modulus} $m \in {\cal M}_r$.
In particular, the query complexity $k$ of the MVC  is  equal to
the number of monomials in the $S$-decoding polynomial, and
 is at most $2^r$ for all  $m\in {\cal M}_r$.
A number $m\in {\cal M}_r$ has been called {\em good}
if  an $S$-decoding polynomial with
$k<2^r$ monomials exists when   $m$ is used to construct MVC.
For example,  the 3-query LDC of
 \cite{Efr09} was constructed with the good number  $m=7\times 73$.
Itoh and Suzuki \cite{IS08} showed that one can reduce the query complexity
of MVCs via  a composition theorem.
In particular, by using the good numbers
  511 and 2047, they were able to obtain $9\cdot 2^{r-4}$-query LDC
of length $N_r$  for all $r>5 $.
 Chee et al. \cite{CFL13} showed that if
there exist  primes $t,p_1,p_2$ such that $m=2^t-1=p_1p_2$,
then $m$
must be good. They determined 50 new  good numbers of the above form and then significantly reduced the query complexity of MVCs.

 Since \cite{Efr09,IS08,CFL13},  the work of {finding}  good numbers  has become
  interesting.
However, the study of  \cite{Efr09,IS08,CFL13} was limited to  good numbers of the form   $m=p_1p_2\in {\cal M}_2$.
When $\max\{\alpha_1,\alpha_2\}>1$,  it is  not known how to decide a number of
 the form $m=p_1^{\alpha_1}p_2^{\alpha_2}\in {\cal M}_2$ is good except using
 the very expensive computer search.
  In this paper, we shall provide two methods for obtaining new good numbers in ${\cal M}_2$:
\begin{itemize}
\item  If    $m_1=p_1^{\alpha_1}p_2^{\alpha_2}\in {\cal M}_2$  is good and $m_2
 =p_1^{\beta_1}p_2^{\beta_2}\in {\cal M}_2$ is  a multiple of $m_1$,
 then   $m_2$ must be good as well;
\item  If    $m_2=p_1^{\beta_1}p_2^{\beta_2}\in {\cal M}_2$ is good,
  and  there is an   $S$-decoding polynomial of the form
 $P(X)=X^u+aX^v+b$ for $m_2$ such that $\gcd(u,v,m_2)=p_1^{\omega_1}p_2^{\omega_2}$, then
 $m_1=m_2/(p_1^{\omega_1}p_2^{\omega_2})$ must be good as well.
\end{itemize}

\section{Preliminaries}

We denote by $\mathbb{Z}$ and $\mathbb{Z}^+$
 the set of integers and  positive integers, respectively.
For any  $n\in \mathbb{Z}^+$, we denote $[n]=\{1,2,\ldots,n\}$.
For any  $m\in \mathbb{Z}^+$, we denote by $\mathbb{Z}_m$
the set of integers modulo $m$ and denote by
$\mathbb{Z}_m^*$ the multiplicative group of integers
modulo $m$.
When $m$ is odd, we have that $2\in \mathbb{Z}_m^*$ and denote by
${\rm ord}_m(2)$ the multiplicative order of $2$ in
$\mathbb{Z}_m^*$.
For a prime power $q$, we denote by
$\mathbb{F}_q$ the finite field of $q$ elements and denote by
$\mathbb{F}_q^*$ the multiplicative group of $\mathbb{F}_q$.
 For any $z\in \mathbb{F}_q^*$, we denote
 by ${\rm ord}_q(z)$ the multiplicative  order of $z$.
For any two vectors $x=(x_1,\ldots,x_n),y=(y_1,\ldots,y_n)$, we denote
by $d_H(x,y)=\{i: i\in [n], x_i\neq y_i\}$ the
{\em Hamming  distance} between $x$ and $y$.
For any   $x,y\in \mathbb{Z}_m^h$, we denote by
$\langle x,y \rangle_m=\sum_{i=1}^n x_iy_i \bmod m$
the dot product  of $x$ and $y$.
If the components of a vector $y$ are labeled by a set $V$, then
for every $v\in V$ we denote by $y[v]$ the $v$th component of $y$.

\begin{definition}
\label{def:ldc}
{\bf (locally decodable code)}
Let $k,n,N\in \mathbb{Z}^+$   and let
$0\leq \delta,\epsilon \leq 1$.
 A code
 $C:\mathbb{F}_q^n\longrightarrow \mathbb{F}_q^N$ is said to be
  {\rm $(k, \delta, \varepsilon)$-locally decodable}  if
  there exist randomized decoding algorithms
    $D_1,D_2,\ldots,D_n$ such that:
    \begin{itemize}
    \item For any   $x\in \mathbb{F}_q^n$,  any  $y\in \mathbb{F}_q^N$
    such that $d_H(C(x),y)\leq \delta N$ and any $i\in[n]$,
  $\Pr[D_i(y)=x_i]\geq 1-\epsilon$.

   \item The algorithm $D_i$  makes at most $k$ queries to $y$.
  \end{itemize}
\end{definition}
The numbers  $k$ and $N$ are called the {\em query complexity}
and the {\em  length} of $C$, respectively.
They are usually considered as
  functions of $n$, the {\em message length}, and measure  the
efficiency of $C$. Ideally, we would like $k$ and $N$ to be as small as possible.

 Efremenko
\cite{Efr09} proposed a construction of LDCs, which is based on two fundamental building blocks:
{\em $S$-matching family} and {\em $S$-decoding polynomial}.

\begin{definition}{\bf ($S$-matching family)}
Let $m,h,n\in \mathbb{Z}^+$ and let  $S\subseteq \mathbb{Z}_m\setminus\{0\}$.
A set ${\cal U}=\{u_i\}_{i=1}^{n}\subseteq \mathbb{Z}_m^h$ is said to be
an {\rm $S$-matching family} if
\begin{itemize}
\item $\langle u_i,u_i\rangle_m =0$ for every $i\in [n]$,
\item $\langle u_i,u_j\rangle_m \in S$ for all  $i,j\in[n]$ such that $i\neq j$.
\end{itemize}
\end{definition}

\begin{definition}{\rm ({\bf $S$-decoding polynomial})}
 Let $m\in \mathbb{Z}^+$ be  odd.
 Let $t={\rm ord}_m(2)$ and let $\gamma\in \mathbb{F}_{2^t}^*$ be a
 primitive $m${\rm th} root of unity.
A polynomial $P(X)\in \mathbb{F}_{2^t}[X]$ is said to be
 an {\rm $S$-decoding polynomial} if
\begin{itemize}
\item $P(\gamma^s)=0$ for every $s\in S$,
\item $P(\gamma^0)=1$.
\end{itemize}

\end{definition}

  Given an $S$-matching family ${\cal U}=\{u_i\}_{i=1}^n
\subseteq \mathbb{Z}_m^h$  and an $S$-decoding polynomial
$P(X)=a_0+a_1X^{b_1}+\cdots + a_{k-1} X^{b_{k-1}}\in \mathbb{F}_{2^t}[X]$,
Efremenko's LDC    can be described
by {\bf Figure.   1}.

\vspace{3mm}
\hrule
\vspace{2mm}
\noindent
{\bf Encoding:}
This algorithm encodes any message
  $x=(x_1,\ldots,x_n)\in \mathbb{F}_{2^t}^n$  as a codeword $C(x)\in \mathbb{F}_{2^t}^{m^h}$ such that:
\begin{itemize}
\item the $m^h$ components  of $C(x)$ are
 labeled by the $m^h$ elements of $\mathbb{Z}_m^h$   respectively;
 and
 \item  for every $v\in \mathbb{Z}_m^h$, the $v$th component is computed as $C(x)[v]=
 \sum_{j=1}^n x_j  \cdot \gamma^{\langle u_j,v \rangle_m}$.
\end{itemize}
\noindent {\bf Decoding:}
This algorithm takes a word $y\in\mathbb{F}_{2^t}^{m^h}$
and an integer $i\in[n]$ as input. It recovers $x_i$ as follows:
\begin{itemize}
\item
 choose a vector  $v\in \mathbb{Z}_m^h$ uniformly and at random.
 \item
 output $\gamma^{-\langle u_i,v\rangle_m}\cdot  (a_0 \cdot  y[v]+\sum_{\ell=1}^{k-1}
 a_\ell \cdot  y[v+b_\ell u_i])$.
\end{itemize}

\vspace{1mm}

\hrule

\vspace{1mm}

\begin{center}
{\bf Figure 1.} Efremenko's Construction
\end{center}

 Efremenko's construction
gives a linear $(k,\delta,k\delta)$-LDC that  encodes
messages of length $n$
to codewords of length $N=m^h$.
 When $N$ is fixed, the larger the $n$ is, the more efficient the
 $C$ is. Efremenko \cite{Efr09} and several later works \cite{IS08,CFL13}
choose $S$ as the   canonical set in $\mathbb{Z}_m$.
For any  $m=p_1^{\alpha_1}p_2^{\alpha_2}\cdots p_r^{\alpha_r}\in {\cal M}_r$,
the {\em canonical set}   in  $\mathbb{Z}_m$ is defined as
 $$S_m=\big\{s_\sigma: \sigma\in \{0,1\}^r\setminus \{0^r\}, s_\sigma\equiv \sigma_i (\bmod~ p_i^{\alpha_i}), \forall i\in[r] \big\}.$$
  For example, $S_{15}=\{1,6,10\}$.
Efremenko \cite{Efr09} observed that
Grolmusz's set system  \cite{Gro00} gives  a direct construction of $S_m$-matching families.
\begin{lemma}
\label{lem:mf}
{\bf (\cite{Gro00,Efr09})}
For  any $m\in {\cal M}_r (r\geq 2)$ and integer $h>0$,
 there is an $S_m$-matching family  ${\cal U}=\{u_i\}_{i=1}^{n}\subseteq
\mathbb{Z}_m^h$  of size $n\geq\exp (c(\log h)^r/(\log\log h)^{r-1})$, where $c$ is a constant that only depends on $m$.
\end{lemma}
In  particular, the $n$ takes the form of $\ell^\ell$ for an integer
$\ell>0$ and $h$ is determined by both $m$, $\ell$, and
the weak representation of the function ${\sf OR}_\ell$ \cite{Gro00}.
Efremenko \cite{Efr09} also observed that the polynomial $P(X)=
\prod_{s\in S_m} (X-\gamma^s)/(1-\gamma^s)$ is an
$S_m$-decoding polynomial with  $k\leq 2^r$ monomials.
\begin{lemma}
\label{lem:dp}
{\bf (\cite{Efr09})}
For any     $m\in{\cal M}_r (r\geq 2)$,
 there is an $S_m$-decoding polynomial  with at most  $  2^r$
 monomials.
\end{lemma}

{Lemmas} \ref{lem:mf} and \ref{lem:dp} yield    LDCs of subexponential length.
\begin{theorem}
\label{thm:ldc}
{\bf(\cite{Efr09})}
For every integer $r\geq 2$, there is a $(k,\delta,k\delta)$-LDC
of
query complexity $k\leq 2^r$
and   length
$N_r$.
\end{theorem}
For every integer $r\geq 2$, Theorem  \ref{thm:ldc} gives
an infinite family of LDCs, each based on a number
$m\in {\cal M}_r$.
Different  $m\in {\cal M}_r$ may  give LDCs of different query complexity.
For example,
  $m=7\times 73$ gives a code of query complexity 3 \cite{Efr09},
  while  $m=3\times 5$ is only able to give a code of query complexity
4 \cite{IS08}.
A number of the form $m=p_1p_2$
has been called {\em good}  in  \cite{IS08,CFL13} if it is able to result in
an LDC of query complexity $<4$.
By using  the good numbers $511$
and $2047$,  Itoh and Suzuki \cite{IS08} concluded that
for any    $r>5$, the query complexity $2^r$ of the LDCs in Theorem  \ref{thm:ldc}
 can be reduced to  $9\cdot 2^{r-4} $.
{On the other hand, for $r=2,3,4$ and $5$,
the best  decoding algorithms to date for the LDCs in Theorem  \ref{thm:ldc}
have query complexity  3, 8, 9, and 24, respectively.}
Chee et al. \cite{CFL13}  showed
  Mersenne numbers of the form $p_1p_2$ are good.
With  {infinitely many} such   good numbers,  Chee et al. \cite{CFL13}
can further reduce  the query complexity to $3^{r/2}$.

\section{Good Numbers  of the Form $\lowercase{p}_1^{\alpha_1} \lowercase{p}_2^{\alpha_2}$}

\label{sec:two}
   Let $m= p_1^{\alpha_1}p_2^{\alpha_2}\in \mathcal{M}_2$.
Let $t={\rm ord}_m(2)$   and let $\gamma \in \mathbb{F}_{2^t}^*$ be a primitive $m{\rm th}$ root of unity.
Lemma \ref{lem:dp} shows that there is an $S_m$-decoding polynomial $P(X)$
with $k\leq 4$ monomials.
In this section, we will    establish several sufficient and necessary conditions for  a number $m$ to be good.

\begin{lemma}
\label{lem:sdp}
 Let $m= p_1^{\alpha_1}p_2^{\alpha_2}\in {\cal M}_2$. Then  any $S_m$-decoding polynomial has $\geq 3$
monomials.
\end{lemma}

\begin{proof}
If
 $P(X)=aX^u\in \mathbb{F}_{2^t}[X]$ is an
  $S_m$-decoding polynomial, then
 $a=P(1)=1$ and $a=\gamma^{-u}P(\gamma)=0$, which give  a contradiction.
If  $P(X)=aX^u+bX^v\in \mathbb{F}_{2^t}[X]$ is an $S_m$-decoding
 polynomial with 2 monomials, then   $ab\neq 0$ and
 \begin{align}
\label{eqn:e2}a\gamma^{us_{01}}+b\gamma^{vs_{01}} &= P(\gamma^{s_{01}})=0,  \\
\label{eqn:e3} a\gamma^{us_{10}}+b\gamma^{vs_{10}} &= P(\gamma^{s_{10}})=0, \\
\label{eqn:e4} a +b &= P(1)\hspace{4.4mm}=1.
\end{align}
Equations (\ref{eqn:e2}) and (\ref{eqn:e3}) imply that
$b/a=\gamma^{(u-v)s_{01}}=\gamma^{(u-v)s_{10}}$.
As ${\rm ord}_{2^t}(\gamma)=m$, we must have that
\begin{equation}
\label{eqn:e5}
(u-v)(s_{01}-s_{10})\equiv 0 ~(\bmod~m).
\end{equation}
Note that  $\gcd(s_{01}-s_{10}, m)=1$.
Equation (\ref{eqn:e5}) implies
$
u\equiv v ~(\bmod ~m).
$
It follows that $b/a=\gamma^{(u-v)s_{01}}=1$ and thus $a+b=0$, which contradicts to
  (\ref{eqn:e4}).
\end{proof}

Let $\mathbb{M}_2$
be the set of good numbers in ${\cal M}_2$.
 The following  lemmas characterize   $\mathbb{M}_2$.

\begin{lemma}
\label{lem:chrz1}
 Let $m=p_1^{\alpha_1} p_2^{\alpha_2}\in {\cal M}_2$. Then   $m\in \mathbb{M}_2$ if and only if there is
a polynomial
   $Q(X)=X^u+aX^v+b \in \mathbb{F}_{2^t}[X]$ that satisfies the following properties
\begin{itemize}
\item[\ding{172}]  $ab\neq 0, $
\item[\ding{173}] $|\{(u\bmod~m), (v\bmod ~m),0\}|=3$, and
\item[\ding{174}]   $Q(\gamma )=Q(\gamma^{s_{01}}) = Q(\gamma^{s_{10}})=0$,
$Q(1) \neq 0$.
\end{itemize}
\end{lemma}

\begin{proof}
If $m\in \mathbb{M}_2$, then  by Lemma \ref{lem:sdp}
 there exists  an $S_m$-decoding polynomial
 \begin{equation}
 \label{eqn:sdp1}
 P(X)=c_1 X^{d_1}+c_2 X^{d_2}+c_3 X^{d_3}\in \mathbb{F}_{2^t}[X]
  \end{equation}
with exactly  3 monomials.
 In particular,  we must have
 \begin{itemize}
 \item[\ding{175}]  $c_1c_2c_3\neq 0$,
 \item[\ding{176}]  $|\{(d_1 \bmod~m),
 (d_2~\bmod~m),(d_3~\bmod~m)\}|=3$, and
  \item[\ding{177}] $P(\gamma^{s_{01}})=P(\gamma^{s_{10}})=P(\gamma)=0, P(1)=1$.
 \end{itemize}
While \ding{175} and \ding{177} are clear from the definition, we show that
\ding{176} is also true.
Assume for contradiction that
 $d_1 \equiv d_2(\bmod~m)$.
 Then
$(\gamma^{s})^{d_1}=(\gamma^s)^{d_2}$
for all $s\in \{s_{01}, s_{10}, 1\}$ and thus
 \begin{align}
\label{eqn:e7} (c_1+c_2)\gamma^{s_{01}d_1}+c_3\gamma^{s_{01} d_3}&=P(\gamma^{s_{01}})=0,  \\
\label{eqn:e8} (c_1+c_2)\gamma^{s_{10}d_1}+c_3\gamma^{s_{10} d_3}&=P(\gamma^{s_{10}})=0, \\
\label{eqn:e9}  c_1+c_2+c_3&=P(1) \hspace{4.4mm}=1.
\end{align}
Due to  (\ref{eqn:e7}) and  (\ref{eqn:e8}), we have that
 $\gamma^{s_{01}(d_3-d_1)}=\gamma^{s_{10}(d_3-d_1)}$ and thus
$d_1\equiv d_3 (\bmod ~m)$.
Consequently, (\ref{eqn:e7}) implies that $c_1+c_2+c_3 =0$, which contradicts to
  (\ref{eqn:e9}).

W.l.o.g., we suppose that $d_1>d_2> d_3$.
   Let $u=d_1-d_3, v=d_2-d_3, a=c_2/c_1$ and  $ b=c_3/c_1$. Then
\begin{equation}
Q(X):=X^u+aX^v+b=\frac{P(X)}{c_1X^{d_3}}.
\end{equation}
The properties \ding{172}, \ding{173}, and \ding{174} trivially follow from
\ding{175}, \ding{176}, and \ding{177}, respectively.

 Conversely, suppose that $Q(X)=X^u+aX^v+b$ is a polynomial that satisfies
  the properties \ding{172}, \ding{173}, and \ding{174}.
   Then $P(X)=Q(X)/Q(1)$ will be an $S_m$-decoding polynomial with exactly {3 monomials}.
 Therefore, $m\in \mathbb{M}_2$.
 \end{proof}

\begin{lemma}
\label{lem:chrz2}
 Let $m=p_1^{\alpha_1} p_2^{\alpha_2}\in {\cal M}_2$.
Then   $m\in \mathbb{M}_2$ if and only if there exist  $u,v\in E:=\{e:p_1^{\alpha_1}\nmid e, p_2^{\alpha_2}\nmid e ,e\in \mathbb{Z}\}$  such that $u\not\equiv v (\bmod ~m)$ and  $\det(A)=0$, where
 \begin{equation}
 \label{eqn:A}
 A=\left(
 \begin{matrix}
  \gamma^{s_{01}u} & \gamma^{s_{01}v} & 1\\
  \gamma^{s_{10}u} & \gamma^{s_{10}v} & 1 \\
   \gamma^{u} & \gamma^{v} & 1 \\
\end{matrix}
\right).
\end{equation}
\end{lemma}

\begin{proof}

 If  $m\in \mathbb{M}_2$, then
   there is
a polynomial
   $Q(X)=X^u+aX^v+b \in \mathbb{F}_{2^t}[X]$   such that
the \ding{172}, \ding{173}, and \ding{174} in Lemma \ref{lem:chrz1} are true.
Due to \ding{173}, we have that $u\not \equiv v (\bmod ~m)$.
On the other hand, \ding{174} is equivalent to
 \begin{align}
\label{eqn:A1ab=0}A
\left(
 \begin{matrix}
1\\
a\\
b\\
\end{matrix}
\right)
&= \left(
\begin{matrix}
 0 \\
 0 \\
 0\\
\end{matrix}
\right),\\
\label{eqn:1+a+bne0}
 1+a+b& \neq 0.
\end{align}
Equation (\ref{eqn:A1ab=0})
requires that $\det(A)=0$.
It remains to show that
$u,v\in E$.
We show that $p_1^{\alpha_1}  \nmid u$.    The proofs for $ p_2^{\alpha_2} \nmid u, p_1^{\alpha_1}  \nmid v,$ and $
 p_2^{\alpha_2} \nmid v$ will be similar and omitted.
Note that
  \begin{equation}
  \label{eqn:det}
  \begin{aligned}
0=\det(A)= &(\gamma^{s_{01}u}+ \gamma^{u})(\gamma^{s_{10}v}+ \gamma^{v})+\\
&(\gamma^{s_{01}v} + \gamma^{v})(\gamma^{s_{10}u} +\gamma^{u})\\
=&((\gamma^{-s_{10}u}+1)(\gamma^{-s_{01}v}+1)+\\
&\hspace{1.2mm}(\gamma^{-s_{01}u}+1)
(\gamma^{-s_{10}v}+1))\gamma^{u+v}.
\end{aligned}
\end{equation}
If $p_1^{\alpha_1}|u$, then  $s_{10}u\equiv 0 ~(\bmod~m)$ and
 $\gamma^{-s_{10}u}+1=0$.
Equation (\ref{eqn:det}) would imply
 $\gamma^{-s_{01}u}+1 =0$
 or   $\gamma^{-s_{10}v}+1=0$.
 If $\gamma^{-s_{01}u}+1 =0$, then
  $p_2^{\alpha_2}|u$ and thus $m|u$, which would
   contradicts to
\ding{173}. If $\gamma^{-s_{10}v}+1=0$, then
   $p_1^{\alpha_1}|v$ and thus
  $0=Q(\gamma^{s_{10}})=1+a+b$, which contradicts to
  (\ref{eqn:1+a+bne0}).

 Conversely, suppose that $u,v\in E$ are integers such that
 $ u\not \equiv v (\bmod~m)$ and  $\det(A)=0$.
To show that $m\in \mathbb{M}_2$, it suffices to construct   an $S_m$-decoding polynomial
$Q(X)=X^u+aX^v+b\in \mathbb{F}_{2^t}[X]$ such that
\ding{172}, \ding{173}, and \ding{174} are satisfied.
First of all, $\det(A)=0$ implies that ${\rm rank}(A)\leq 2$. If ${\rm rank}(A)=1$, then
we must have that
$\gamma^{s_{01}u}=\gamma^{s_{10}u}$. It follows that
  $u\equiv 0 (\bmod ~m)$, which contradicts to
  $u\in E$.
As ${\rm rank}(A)=2$,  the null space of $A$
will be 1-dimensional and spanned by
a nonzero vector $c=(c_1,c_2,c_3)^T$.
Below we shall see that $c_i\neq 0$ for all $ i\in[3]$.
If $c_1=0$, then
\begin{equation}
 \left(
 \begin{matrix}
  \gamma^{s_{01}v} & 1\\
  \gamma^{s_{10}v} & 1 \\
  \gamma^{v} & 1 \\
\end{matrix}
\right)
\left(
 \begin{matrix}
  c_2 \\
c_3\\
\end{matrix}
\right)
= \left(
\begin{matrix}
 0 \\
 0 \\
 0\\
\end{matrix}
\right).
\end{equation}
Then $c_2c_3$ must be nonzero and thus
  $\gamma^{s_{01}v} = \gamma^{s_{10}v}$. The latter equality requires that
 $v\equiv 0 (\bmod~m)$, which contradicts to the fact $v\in E$.
 Hence,  $c_1\neq 0$.
Similarly, we have $c_2\neq 0$ and $c_3\neq 0$.
Let
$R(X)=c_1X^u+c_2X^v+c_3$. Then
    $R(\gamma)=R(\gamma^{s_{01}}
) = R(\gamma^{s_{10}})=0$.
Furthermore, we must have $R(1)\neq 0$.
Otherwise,  $c_3=c_1+c_2$ and
\begin{equation}
 \left(
 \begin{matrix}
  \gamma^{s_{01}u}+1 & \gamma^{s_{01}v}+1\\
  \gamma^{s_{10}u}+1 & \gamma^{s_{10}v}+1 \\
  \gamma^{u}+1 & \gamma^{v}+1 \\
\end{matrix}
\right)
\left(
 \begin{matrix}
  c_1 \\
c_2\\
\end{matrix}
\right)
= \left(
\begin{matrix}
 0 \\
 0 \\
 0\\
\end{matrix}
\right).
\end{equation}
As $c_1c_2\neq 0$, this is possible only if
 \begin{equation}
 \label{eqn:e17}
\frac{\gamma^{s_{01}u}+1}{\gamma^{s_{01}v}+1}=\frac{\gamma^{s_{10}u}+1}{\gamma^{s_{10}v}+1}
=\frac{\gamma^u+1}{\gamma^v+1}.
\end{equation}
{Denote by $\lambda$ the value of the fractions in (\ref{eqn:e17}). Then
\begin{equation}
\label{eqn:ss1}
\hspace{0mm}\begin{split}
 \frac{\gamma^{s_{10}u}}{\gamma^{s_{10}v}}\cdot
 \underbrace{~\frac{\gamma^{s_{01}u}+1}{\gamma^{s_{01}v}+1}~}_{\lambda}
 &=
\frac{\gamma^{s_{10}u}+\gamma^u}{\gamma^{s_{10}v}+\gamma^v}\\
 &=
\frac{\gamma^{s_{10}u}+1+\gamma^u+1}{\gamma^{s_{10}v}+1+\gamma^v+1}\\
 &=
\frac{\lambda(\gamma^{s_{10}v}+1)+\lambda(\gamma^v+1)}{\gamma^{s_{10}v}+1+\gamma^v+1}\\
 &=\lambda,
\end{split}
 \end{equation}
 where the first equality is based on the fact that $s_{01}+s_{10}\equiv 1 (\bmod ~m)$ and the second equality is  true because we are working over
 a finite field of characteristic 2.
It follows from     (\ref{eqn:ss1})  that   $\gamma^{s_{10}(u-v)}=1$.
Therefore,  we must have that    $u\equiv v (\bmod~p_1^{\alpha_1})$.
Similarly, we have that
$u\equiv v (\bmod~p_2^{\alpha_2})$.
Based on the two congruences,   we have that
$u\equiv v (\bmod~m)$, which gives a contradiction.}
 Hence,  $R(1)\neq 0$ and  $Q(X):=R(X)/c_1$ is a polynomial satisfying
 \ding{172},  \ding{173}, and  \ding{174}.
\end{proof}

 \begin{lemma}
\label{lem:chrz3}
 Let $m=p_1^{\alpha_1} p_2^{\alpha_2}\in {\cal M}_2$.
Let $t={\rm ord}_m(2)$  and let $\gamma \in \mathbb{F}_{2^t}^*$ be a primitive $m${\rm th} root of unity.
Let $$\tau(z_1,z_2)=\frac{z_1+z_2}{z_1z_2+z_2}.$$
Then
 $m \in \mathbb{M}_2$ if and only if
   $\tau$ is not injective on
   ${\cal D}=\{(z_1,z_2)\in
(\mathbb{F}_{2^t}^*\setminus\{1\})^2: {\rm ord}_{2^t} (z_1)| p_1^{\alpha_1},
 {\rm ord}_{2^t}(z_2)|p_2^{\alpha_2}\}$.
\end{lemma}

\begin{proof}
If  $m \in \mathbb{M}_2$, then by
Lemma \ref{lem:chrz2}   there  exist
 $ u,v\in E$  such that $u\not\equiv v(\bmod~m)$ and
  $\det(A)=0$, where $A$ is defined by (\ref{eqn:A}).
 Note that $\det(A)=0$ requires that
 $$
 \frac{\gamma^{s_{10}u}+\gamma^{s_{01}u}}{\gamma^u+\gamma^{s_{01}u}}=
  \frac{\gamma^{s_{10}v}+\gamma^{s_{01}v}}{\gamma^v+\gamma^{s_{01}v}}.
 $$
Clearly,
 $(\gamma^{s_{10}u},\gamma^{s_{01}u})$ and $
 (\gamma^{s_{10}v},\gamma^{s_{01}v})$ are two distinct
 elements of $\cal D$ and
 $
\tau(\gamma^{s_{10}u},\gamma^{s_{01}u} )
=
\tau(\gamma^{s_{10}v},\gamma^{s_{01}v} ).
$
Hence, $\tau$ is not injective on $\cal D$.

Conversely, suppose that $\tau(z_1,z_2)=\tau(z_1^\prime,z_2^\prime)$
for two distinct elements
$(z_1,z_2), (z_1^\prime,z_2^\prime)\in {\cal D} $.
To show that $m\in \mathbb{M}_2$, by Lemma \ref{lem:chrz2}  it suffices to
find
 $ u,v\in E$  such that $u\not\equiv v(\bmod~m)$ and
  $\det(A)=0$.
  Suppose that
  $${\rm ord}_{2^t}(z_1)=p_1^{i_1},  {\rm ord}_{2^t}(z_2)=p_2^{j_1},
   $$$${\rm ord}_{2^t}(z_1^\prime)=p_1^{i_2},
{\rm ord}_{2^t}(z_2^\prime)=p_2^{j_2}$$
 for  $i_1,i_2\in [\alpha_1]$ and $j_1,j_2\in [\alpha_2]$.
 Then there exist integers
 $u_1, u_2, v_1,v_2$, where $p_1\nmid u_1, v_1$ and $ p_2\nmid u_2, v_2$,  such that
 $$z_1=(\gamma^{s_{10}p_1^{\alpha_1-i_1}})^{u_1},
 z_2=(\gamma^{s_{01}p_2^{\alpha_2-j_1}})^{u_2},
 $$$$
  z_1'=(\gamma^{s_{10}p_1^{\alpha_1-i_2}})^{v_1}, z_2'=(\gamma^{s_{01}p_2^{\alpha_2-j_2}})^{v_2}. $$
By Chinese remainder theorem,  there exist   $u, v$ such that:
\begin{equation*}
\left\{
  \begin{array}{ll}
    u\equiv p_1^{\alpha_1-i_1}u_1 (\bmod~p_1^{\alpha_1}),   \\
    u\equiv p_2^{\alpha_2-j_1}u_2 (\bmod~p_2^{\alpha_2}),
  \end{array}
\right.
\left\{
  \begin{array}{ll}
    v\equiv p_1^{\alpha_1-i_2}v_1 (\bmod~p_1^{\alpha_1}),   \\
v\equiv p_2^{\alpha_2-j_2}v_2 (\bmod~p_2^{\alpha_2}).
  \end{array}
\right.
\end{equation*}
In particular,  $u, v \in E$ and $u\not\equiv v(\bmod~m)$ (o.w., we will have
 $(z_1,z_2)=(z_1',z_2')$). Furthermore,
$z_1=\gamma^{s_{10}u}, z_2=\gamma^{s_{01}u}, z_1'=\gamma^{s_{10}v}$ and $z_2'=\gamma^{s_{01}v}.$
 Since
 $\tau(z_1,z_2)=\tau(z_1^\prime,z_2^\prime)$, we must   have that  $\det(A)=0$
 due to (\ref{eqn:det}).
 \end{proof}

Let  $\rho(z_1,z_2)=\tau(z_1,z_2)-1=
(1+z_2^{-1})(1+z_1^{-1})^{-1}.$
Then Lemma  \ref{lem:chrz3} gives the following theorem.
 \begin{theorem}
 \label{the:jug}
 Let $m=p_1^{\alpha_1} p_2^{\alpha_2}\in {\cal M}_2$.
Let $t={\rm ord}_m(2)$   and let $\gamma \in \mathbb{F}_{2^t}^*$ be a primitive $m{\rm th}$ root of unity.
Then
 $m \in \mathbb{M}_2$ if and only if
$\rho$ is not injective on $\cal D$.
 \end{theorem}
Theorem \ref{the:jug} gives a characterization of the   good numbers in ${\cal M}_2$.
We say that  $(u,v)\in E^2$ form a {\em collision} for $m$ if
\begin{itemize}
\item
$\rho(\gamma^{s_{10}u},\gamma^{s_{01}u})=\rho(\gamma^{s_{10}v},\gamma^{s_{01}v})$, and
\item
$u\not\equiv v(\bmod~m)$.
\end{itemize}
The proof of Lemma  \ref{lem:chrz3} shows that  $m\in \mathbb{M}_2$
 if and only if there is a collision  $(u, v)\in E^2$ for $m$.

\section{Implications between Good Numbers}

In this section, we show the implications between good numbers in
${\cal M}_2$, which    allows us to
{construct} new good numbers from old.

\begin{lemma}
\label{the:gam}
 Let $m_1=p_1^{\alpha_1} p_2^{\alpha_2}, m_2=p_1^{\beta_1} p_2^{\beta_2}
 \in {\cal M}_2$.
Let $t_i={\rm ord}_{m_i}(2)$ and let $\gamma_i \in \mathbb{F}_{2^{t_i}}^*$ be a primitive $m_i{\rm th}$ root of unity for $i=1,2$.
If $m_1|m_2$, then there is an integer  $\sigma\in \mathbb{Z}_{m_1}^*$ such that
  $\gamma_1=\gamma_2^{\sigma m_2/m_1}$.
\end{lemma}

\begin{proof}
As  $m_1|m_2$ and $m_2|(2^{t_2}-1)$, we must have that
$t_1|t_2$.
Then   $\mathbb{F}_{2^{t_1}}$ is  a subfield of $\mathbb{F}_{2^{t_2}}$.
{Note that  $\gamma_1\in \mathbb{F}_{2^{t_1}}\subseteq \mathbb{F}_{2^{t_2}}$
and $\gamma_2^{m_2/m_1} \in \mathbb{F}_{2^{t_2}}$ are elements of the same finite field
and have the same multiplicative order (i.e., $m_1$).
Both
 $\langle \gamma_1\rangle$ and $\langle
\gamma_2^{m_2/m_1}\rangle$ are subgroups   of
$\mathbb{F}_{2^{t_2}}^*$ of order $m_1$. As $\mathbb{F}_{2^{t_2}}^*$ has a unique subgroup of
order $m_1$, it must be the case that  $\langle \gamma_1\rangle=\langle
\gamma_2^{m_2/m_1}\rangle$.} Hence, there is an integer
$\sigma\in \mathbb{Z}_{m_1}^*$ such that
$ \gamma_1=\gamma_2^{\sigma m_2/m_1}$.
\end{proof}

\begin{theorem}
\label{thm:im1}
 Let $m_1=p_1^{\alpha_1} p_2^{\alpha_2},  m_2=p_1^{\beta_1} p_2^{\beta_2}\in {\cal M}_2$.
 If $m_1\in \mathbb{M}_2$ and $m_1|m_2$, then $m_2\in \mathbb{M}_2$.
\end{theorem}
\begin{proof}
For   $i\in \{1,2\}$,  let  $S_{m_i}=\{s^i_{01},s^i_{10},  1 \}$,  let $t_i={\rm ord}_{m_i}(2)$, and let $\gamma_i \in \mathbb{F}_{2^{t_i}}^*$ be of order  $m_i$.
 Let $E_1=\{e:p_1^{\alpha_1}\nmid e, p_2^{\alpha_2}\nmid e, e\in \mathbb{Z}\}$ and
$ E_2=\{e:p_1^{\beta_1}\nmid e, p_2^{\beta_2}\nmid e, e\in \mathbb{Z}\}$.

If $m_1\in \mathbb{M}_2$, then there is a collision
  $(u_1,v_1)\in E^2$ such that $u_1\not\equiv v_1 (\bmod~m_1)$ and
\begin{equation}
\label{eqn:col}
\frac{\gamma_1^{-s^1_{01}u_1}+1}{\gamma_1^{-s^1_{10}u_1}+1}=\frac{\gamma_1^{-s^1_{01}v_1}+1}
 {\gamma_1^{-s^1_{10}v_1}+1}.
\end{equation}
{As per}  Lemma \ref{the:gam}, let
$\sigma\in \mathbb{Z}_{m_1}^*$  be an integer such that
  $\gamma_1=\gamma_2^{\sigma m_2/m_1}$.
Then  (\ref{eqn:col}) is
\begin{equation}
\label{eqn:col1}
\frac{\gamma_2^{-s^1_{01}u_1 \sigma m_2/m_1}+1}{\gamma_2^{-s^1_{10}u_1 \sigma m_2/m_1}+1}
=
\frac{\gamma_2^{-s^1_{01}v_1 \sigma m_2/m_1}+1}{\gamma_2^{-s^1_{10}v_1 \sigma m_2/m_1}+1}.
\end{equation}
We claim that  if there exist integers $u_2, v_2$ such that
$$
{\rm (i)}\left\{
  \begin{array}{ll}
 s^1_{01}u_1 \sigma m_2/m_1  \equiv  s^2_{01}u_2
(\bmod~ m_2),\\
  s^1_{10}u_1 \sigma m_2/m_1  \equiv  s^2_{10}u_2
(\bmod ~m_2),
  \end{array}
\right.$$$$
\hspace{-1.5mm}{\rm (ii)}
\left\{
  \begin{array}{ll}
  s^1_{01}v_1\sigma m_2/m_1  \equiv  s^2_{01}v_2
(\bmod ~m_2),\\
s^1_{10}v_1 \sigma m_2/m_1  \equiv s^2_{10} v_2
(\bmod~ m_2),
  \end{array}
\right.
$$
then $u_2, v_2 \in E_2$, $u_2\not\equiv v_2 (\bmod~m_2)$ and
\begin{equation}
\label{eqn:col2}
\frac{\gamma_2^{-s^2_{01}u_2}+1}{\gamma_2^{-s^2_{10} u_2}+1}
=\frac{\gamma_2^{-s^2_{01} v_2}+1}{\gamma_2^{-s^2_{10} v_2}+1},
\end{equation}
i.e., $(u_2,v_2)$ form a collision for $m_2$ (and thus
 $m_2\in \mathbb{M}_2$).
Note that (\ref{eqn:col2}) is clear from (i), (ii) and
(\ref{eqn:col1}).
We need to show that
$u_2, v_2 \in E_2$ and $u_2\not\equiv v_2 (\bmod~m_2)$.
If $p_1^{\beta_1}|u_2$, then we will have that
$m_2|s^2_{10}u_2$.  The second congruence of (i) would imply that
 $p_1^{\alpha_1}|u_1\sigma$, which contradicts to   $u_1\in E_1$ and $\sigma\in \mathbb{Z}_{m_1}^*$.
Similarly, we have $p_2^{\beta_2}\nmid u_2, p_1^{\beta_1}\nmid v_2$
and $p_2^{\beta_2}\nmid v_2$. Hence,   $u_2, v_2\in E_2$.
If $u_2\equiv v_2 (\bmod~m_2)$, then the first congruences of (i) and (ii)
would imply   that
$s^1_{01}\sigma(u_1-v_1)   \equiv 0 (\bmod~m_1) $, which
requires that  $u_1\equiv v_1(\bmod ~p_2^{\alpha_2})$.
Similarly, the second congruences of (i) and (ii)
would imply  $u_1\equiv v_1(\bmod ~p_1^{\alpha_1})$.
It follows that $u_1\equiv v_1 (\bmod ~m_1)$, which is a contradiction.

It remains to show  the existence of
integers $u_2$ and $v_2$ that satisfy (i) and (ii).
We show that existence of $u_2$. The existence of $v_2$
is similar.
Due to Chinese remainder theorem, the first congruence of
(i) is equivalent to
\begin{equation}
\label{eqn:cgr3}
\left\{
  \begin{array}{ll}
 s^1_{01} u_1\sigma m_2/m_1 \equiv s^2_{01} u_2 (\bmod~ p_1^{\beta_1}),
\\
 s^1_{01}u_1\sigma m_2/m_1 \equiv s^2_{01} u_2 (\bmod~ p_2^{\beta_2}),
  \end{array}
\right.
\end{equation}
Note that the first congruence of (\ref{eqn:cgr3}) is always true.
{On the other hand, as
$s^2_{01}\equiv 1 (\bmod~p_2^{\beta_2})$, the first congruence of (i) must be  equivalent to
 \begin{equation}
 \label{eqn:s1}
 u_2\equiv  s^1_{01} u_1\sigma m_2/m_1   (\bmod~ p_2^{\beta_2}).
 \end{equation}
Similarly,   the second congruence of (i) is equivalent to
 \begin{equation}
  \label{eqn:s2}
 u_2\equiv  s^1_{10} u_1\sigma m_2/m_1   (\bmod~ p_1^{\beta_1}).
 \end{equation}
 }
Therefore, (i) is equivalent to the system formed by
(\ref{eqn:s1}) and (\ref{eqn:s2}).
The existence of
$u_2$ {is}  an easy consequence of   the
Chinese remainder theorem.
\end{proof}

\begin{theorem}
\label{thm:im2}
Let $m_2=p_1^{\beta_1} p_2^{\beta_2}\in {\cal M}_2$.
Suppose that $m_2\in \mathbb{M}_2$  and $(u,v)=(p_1^{i_1}p_2^{i_2}\sigma_1, p_1^{j_1}p_2^{j_2}\sigma_2)$
is a collision for $m_2$, where $\sigma_1,\sigma_2\in \mathbb{Z}_{m_2}^*$.
Let $\omega_1=\min\{i_1,j_1\}$ and $\omega_2=\min\{i_2, j_2\}$.
Then   $m_1:=m_2/(p_1^{\omega_1}p_2^{\omega_2})$ belongs to $\mathbb{M}_2$.
\end{theorem}
\begin{proof}
 For $i=1,2$, let   $S_{m_i}=\{s^i_{01},s^i_{10}, 1\}$,
 let $t_i={\rm ord}_{m_i}(2)$ and
 let    $\gamma_i \in \mathbb{F}_{2^{t_i}}^*$ be of order
 $m_i$. Let
$ E_1=\{e:p_1^{\beta_1-\omega_1}\nmid e, p_2^{\beta_2-\omega_2}\nmid e, e\in \mathbb{Z}\}$ and $E_2=\{e:p_1^{\beta_1}\nmid e,
p_2^{\beta_2}\nmid e, e\in \mathbb{Z}\}$.
To show that $m_1\in \mathbb{M}_2$,  it suffices to find
two integers $u_1,v_1\in E_1$ such that $u_1\not\equiv v_1(\bmod~m_1)$ and
\begin{equation}
\label{eqn:trans}
\frac{\gamma_1^{-s^1_{01}u_1}+1}{\gamma_1^{-s^1_{10}u_1}+1}
=\frac{\gamma_1^{-s^1_{01}v_1}+1}{\gamma_1^{-s^1_{10}v_1}+1}.
\end{equation}
As per Lemma \ref{the:gam}, there is an integer
 $\sigma \in \mathbb{Z}_{m_2}^*$ such that
$\gamma_1=\gamma_2^{ p_1^{\omega_1}p_2^{\omega_2}\sigma}$. Then (\ref{eqn:trans})
is
\begin{equation}
\label{eqn:col3}
 \frac{\gamma_2^{-s^1_{01}u_1p_1^{\omega_1}p_2^{\omega_2}\sigma }+1}{\gamma_2^{-s^1_{10}u_1p_1^{\omega_1}p_2^{\omega_2}\sigma }+1}
=\frac{\gamma_2^{-s^1_{01}v_1p_1^{\omega_1}p_2^{\omega_2}\sigma }+1}{\gamma_2^{-s^1_{10}v_1p_1^{\omega_1}p_2^{\omega_2}\sigma }+1}.
\end{equation}
As   $(p_1^{i_1}p_2^{i_2}\sigma_1, p_1^{j_1}p_2^{j_2}\sigma_2)\in E_2^2$ is a
collision for $m_2$, we have
\begin{equation}
\label{eqn:col4}
\frac{\gamma_2^{-s^2_{01}p_1^{i_1}p_2^{i_2}\sigma_1}+1}{\gamma_2^{-s^2_{10}p_1^{i_1}p_2^{i_2}\sigma_1}+1}
=\frac{\gamma_2^{-s^2_{01}p_1^{j_1}p_2^{j_2}\sigma_2 }+1}{\gamma_2^{-s^2_{10} p_1^{j_1}p_2^{j_2}\sigma_2}+1}.
\end{equation}
We claim that  if there exist integers $u_1, v_1$ such that
$$
{\rm (i)}\left\{
  \begin{array}{ll}
  s^1_{01}u_1p_1^{\omega_1}p_2^{\omega_2}\sigma  \equiv  s^2_{01}p_1^{i_1}p_2^{i_2}\sigma_1
 (\bmod~ m_2),\\
  s^1_{10}u_1p_1^{\omega_1}p_2^{\omega_2}\sigma \equiv
 s^2_{10}p_1^{i_1}p_2^{i_2}\sigma_1
(\bmod~m_2),
  \end{array}
\right.
$$$$
\hspace{-1mm}{\rm (ii)}
\left\{
  \begin{array}{ll}
  s^1_{01}v_1p_1^{\omega_1}p_2^{\omega_2}\sigma  \equiv  s^2_{01}p_1^{j_1}p_2^{j_2}\sigma_2
 (\bmod~ m_2),\\
  s^1_{10}v_1p_1^{\omega_1}p_2^{\omega_2}\sigma \equiv
 s^2_{10}p_1^{j_1}p_2^{j_2}\sigma_2
(\bmod~m_2),
  \end{array}
\right.
$$
then  $u_1,v_1 \in E_1$, $u_1 \not\equiv v_1 (\bmod~m_1)$ and
(\ref{eqn:col3})  holds.
Note that (\ref{eqn:col3}) is clear from (i), (ii) and
(\ref{eqn:col4}).
If $p_1^{\beta_1-\omega_1}|u_1$, then
 the second congruence of (i) would imply that
$p_1^{\beta_1}|p_1^{i_1}p_2^{i_2}\sigma_1 $, which contradicts to the fact that
$p_1^{i_1}p_2^{i_2}\sigma_1\in E_2$.
Similarly, we can show that $p_2^{\beta_2-\omega_2}\nmid u_1, p_1^{\beta_1-\omega_1}\nmid v_1$
and $p_2^{\beta_2-\omega_2}\nmid v_1$. Therefore,   $u_1, v_1\in E_1$.
If $u_1\equiv v_1  (\bmod~m_1)$, then the first congruences  of (i) and (ii)
would imply   that
$ s^2_{01}(p_1^{i_1}p_2^{i_2}\sigma_1 -p_1^{j_1}p_2^{j_2}\sigma_2) \equiv 0 (\bmod~m_2) $, which requires that  $p_1^{i_1}p_2^{i_2}\sigma_1 \equiv p_1^{j_1}p_2^{j_2}\sigma_2(\bmod ~p_2^{\beta_2})$.
Similarly, the second congruences of (i) and (ii)
require that $p_1^{i_1}p_2^{i_2}\sigma_1 \equiv p_1^{j_1}p_2^{j_2}\sigma_2(\bmod ~p_1^{\beta_1})$.
It follows that
$p_1^{i_1}p_2^{i_2}\sigma_1  \equiv p_1^{j_1}p_2^{j_2}\sigma_2 (\bmod ~m_2)$, which is a contradiction.

It remains to show  the existence of
 $u_1$ and $v_1$ that satisfy (i) and (ii).
We show the  existence of $u_1$. The existence of $v_1$
is similar and omitted.
As $\omega_1\leq  i_1\leq \beta_1 $ and $\omega_2\leq  i_2\leq \beta_2 $, the first
congruence  in
(i) is equivalent to
\begin{equation}
\label{eqn:cgr}
\left\{
  \begin{array}{ll}
  s^1_{01}u_1 \sigma  \equiv  s^2_{01}p_1^{i_1-\omega_1}p_2^{i_2-\omega_2}\sigma_1
 (\bmod~p_1^{\beta_1-\omega_1}),\\
  s^1_{01}u_1 \sigma  \equiv  s^2_{01}p_1^{i_1-\omega_1}p_2^{i_2-w_2}\sigma_1
 (\bmod~p_2^{\beta_2-\omega_2}).
  \end{array}
\right.
\end{equation}
Note that the first congruence of (\ref{eqn:cgr}) is always true.
On the other hand, as $p_2\nmid s^1_{01}\sigma$, there is
an integer $t^1_{01}$ such that
$s^1_{01}\sigma t^1_{01}\equiv 1 (\bmod~p_2^{\beta_2-\omega_2})$.
Therefore, the first congruence of (i) will be equivalent to
 \begin{equation}
 \label{eqn:crt1}
 u_1  \equiv  s^2_{01}t^1_{01}p_1^{i_1-\omega_1}p_2^{i_2-\omega_2}\sigma_1
 (\bmod~p_2^{\beta_2-\omega_2}).
 \end{equation}
Similarly, we can show that the second congruence of (i) is equivalent to
 \begin{equation}
  \label{eqn:crt2}
 u_1 \equiv  s^2_{10}t^1_{10}p_1^{i_1-\omega_1}p_2^{i_2-\omega_2}\sigma_1
 (\bmod~p_1^{\beta_1-\omega_1}),
 \end{equation}
where $t^1_{10}$ is an integer such that $s^1_{10}\sigma t^1_{10}\equiv 1(\bmod~p_1^{\beta_1-\omega_1})$.
The existence of
$u_1$ is an easy consequence of  the
Chinese remainder theorem on (\ref{eqn:crt1}) and (\ref{eqn:crt2}).
\end{proof}

\begin{example}
Let $m_2 = 7^2\times 151$. Then  $S_{m_2}=\{s_{01}=1813,s_{10}=5587,s_{11}=1\}$.
 Let $t_2={\rm ord}_{m_2}(2)$
and  let  $\gamma_2 \in \mathbb{F}_{2^{t_2}}^*$ be
a primitive $m_2{\rm th}$ root of unity.
 Then   $(238, 455)$ is a collision for $m_2$.
   Clearly, $\omega_1=1$ and $\omega_2=0$.
   Then $m_1=m_2/7=1057$ must be a good number, which is  $<m_2$.
\end{example}

\section{Conclusion}

In this paper,  we  characterized the good numbers
in ${\cal M}_2$
and   showed two implications between good
numbers in ${\cal M}_2$.
In particular, the second implication requires an additional condition.
 It is an interesting problem to remove the condition.

\newpage

\noindent
{\bf Data Availability Statement}
The data underlying this article are available in the article.

\vspace{5mm}
\noindent
{\bf Funding}
	Natural Science Foundation of Shanghai (21ZR1443000);
Singapore Ministry of Education under Research Grant RG12/19

\vspace{5mm}
\noindent
{\bf Acknowledgments}
The authors would like to thank the anonymous reviewers for
their helpful suggestions.

\nocite{*}

\bibliographystyle{compj}

\end{document}